\documentclass[aps,prc,onecolumn,superscriptaddress]{revtex4}

\usepackage[latin1]{inputenc}
\usepackage{bm}
\usepackage{epsfig}
\usepackage{amsthm}
\usepackage{amsfonts}
\usepackage{float}
\usepackage{amsmath,amssymb}
\usepackage{color}
\usepackage{slashed}
\usepackage{relsize}
\usepackage[toc,page]{appendix} 
\usepackage{ulem}
\usepackage{cancel}
\usepackage{tikz}
\usepackage{hyperref}
\usepackage{fontenc}
\usepackage{graphicx}
\usepackage{dcolumn}
\usepackage{bm}
\usepackage{wrapfig}
\usepackage{makecell}

\newcommand{\be}{\begin{equation}} 
\newcommand{\ee}{\end{equation}} 
\newcommand{\bea}{\begin{eqnarray}} 
\newcommand{\eea}{\end{eqnarray}} 
\newcommand{\nn}{\nonumber} 
\newcommand{\mintedim}[2]{{\int\kern-0.50em\mbox{{\small$\mathop{\frac{\mbox{{\small${\rm d^{#2}}\vect{#1}$}}}{\mbox{{\small$(2\pi)^{#2}$}}}}$}}\ }} 
\newcommand{\inteonedim}[1]{{\int_0^\infty\kern-1em\mbox{{\small${\rm d}{#1}$}}}} 
\newcommand{\vect}[1]{\bm{#1}} 

\newcommand{\kt}{k_{\text{T}}}
\newcommand{\ks}{k_{\text{S}}}
\newcommand{\kr}{k_{\text{R}}}
\newcommand{\ka}{k_{\text{A}}}
\newcommand{\qt}{q_{\text{T}}}
\newcommand{\qs}{q_{\text{S}}}
\newcommand{\alpht}{\alpha_{\text{T}}}
\newcommand{\alphs}{\alpha_{\text{S}}}

\begin{document}
%
\title{
\textcolor{black} {A mathematical model of delay discounting with bi-faceted impulsivity}
}
\author{Shanu Shukla}
\email{shanu.shukla11@gmail.com}
\affiliation{Interdisciplinary Research Team on Internet and Society, Faculty of Social Studies, Masaryk University, Brno, Czech Republic}

\author{Trambak Bhattacharyya}
\email{trambak.bhattacharyya@ujk.edu.pl (corresponding author)}
\affiliation{Institute of Physics, Jan Kochanowski University, Kielce 25-406, Poland}


\begin{abstract} 
Existing mathematical models of delay discounting (e. g. exponential model, hyperbolic model, and those derived from nonextensive statistics) consider impulsivity as a single quantity. However, the present article derives a novel mathematical model of delay discounting considering impulsivity as a multi-faceted quantity. It considers impulsivity to be represented by two positive and fluctuating quantities (e.g. these facets may be trait and state impulsivity). To derive the model, the superstatistics method, which has been used to describe fluctuating physical systems like a thermal plasma, has been adapted. According to the standard practice in behavioural science, we first assume that the total impulsivity is a mere addition of the two facets. However, we also explore the possibility beyond an additive model and conclude that facets of impulsivity may also be combined in a non-additive way. We name this group of models the Extended Effective Exponential Model or E$^3$M. We find a good agreement between our model and experimental data.

\vspace{10pt}
\end{abstract}
\maketitle
%


\section{Introduction}

Delay discounting, a well-explored phenomenon in various fields including behavioural science, economics, and finance, describes the tendency for the subjective value of a reward to decrease as its delivery is postponed. This can be illustrated by a hypothetical scenario where an individual is presented with a choice between receiving \$50 immediately or \$70 after a month. Such a decision involves selecting between a smaller-sooner reward and a larger-later reward, influenced by personal factors like impulsivity and situational factors such as trust. Some individuals may opt for the immediate reward, exhibiting impulsive behaviour according to behavioural analysts \cite{ikuk}. 
This behaviour suggests that a distant reward is perceived as less valuable than an immediate one. Delay discounting has been observed across various species, reward types, and sample populations, as evidenced in studies \cite{mazur,bdm,lempert,west,cajueiro}. 
Researchers have proposed several mathematical models to comprehend this phenomenon, including the exponential, hyperbolic, and
$q$-exponential models of delay discounting \cite{expmodel,rachilinhyp,takahashi1}. These models commonly incorporate impulsivity as a parameter. However,
they also often treat impulsivity as a single quantity, while numerous studies acknowledge impulsivity as
a multi-faceted quantity \cite{impmul1}. 

To account for this complexity, we introduce a mathematical model that considers impulsivity as having two facets, as examined in Refs.~\cite{impbi1,impbi2}. As per the standard practice, we first assume an additive model in which the total quantity is a mere addition of the two facets. While addition offers utility, the facets may not be inherently additive. Evidence for this may be found in Ref.~\cite{stamatesimpul} that outlines that impulsivity is multifaceted, and different facets of impulsivity have differential, sometimes opposing relationships with alcohol-related behaviors. That supports the idea that the effects of impulsivity cannot simply be summed up across all facets.
Even if we assume that the facets of impulsivity are independent, addition is merely only one of the many ways to combine them. Hence, in this article, along with an additive model, we also propose a nonadditive model
and analyze existing datasets obtained from previous experiments.

The structure of this paper is as follows: the subsequent section provides a brief overview of existing mathematical models of delay discounting. Section \ref{model} presents a comprehensive derivation of our proposed model.
In Section \ref{data}, we analyze available datasets and present the findings. Finally, we summarize and conclude our study in Section \ref{summary}.

\section{Existing mathematical models of delay discounting}

In the experiments studying delay discounting, the subjective values of a reward are established by finding indifference points \cite{critchfield2001}. 
In the case of rational agents, indifference points follow an exponential decay with delay. This model is called the exponential model
\cite{expmodel} (EM) and considers an exponential dependence of the subjective value of a reward on delay given by
\be
V(D)=V(0)\exp({-\kappa_0 D}),~~~~~~\text{(--EM--)} 
\label{expmodel}
\ee
where $V(0)$ is the undiscounted value of a reward, and $V(D)<V(0)$ is the discounted value of the reward after a delay $D$. The quantity
$\kappa_0>0$ is described as the impulsivity parameter. However, the exponential model describes the situation when the discount rate is a constant,
a characteristic observed for rational agents. Studies in delay discounting recognize inter-temporal consistency as a characterization of rationality. 
However, rationality of agents has long been a matter of discussion and debate \cite{kahneman} in behavioural science. For non-rational agents, one needs to look 
beyond the exponential model and previous researchers utilized the hyperbolic model \cite{rachilinhyp}. 
%
%


However, the hyperbolic model can not distinguish between inter-temporal inconsistency and impulsivity, and hence Takahashi \cite{takahashi1} proposed the following model, also utilized in Ref.~\cite{cajueiro}, containing the $q$-exponential function
\bea
V(D) 
&=& \frac{V(0)} {\left(1+(1-q)\kappa_1 D\right)^{\frac{1}{1-q}}}. 
\label{Takahashimodel}
\eea
The $q$-exponential model has been used in many other behavioral studies \cite{takahashi2,takahashiqdt,lempert,fpsy1,fph1,ejmbe}.


\subsubsection{Revisiting the effective Exponential Model}
\label{seceem}

Models given by Eqs.~\eqref{expmodel}-\eqref{Takahashimodel} (or the ones equivalent to them) follow a phenomenological approach. However, we are interested in another formulation, called superstatistics, whose application in social systems was proposed recently in Ref.~\cite{tbssrpphysica}. This approach considers fluctuations in the scale parameter in an exponential model. For the physical systems, this leads to temperature fluctuation, and for the delay discounting model, this implies impulsivity to be considered as a random variable. Interestingly, superstatistics was first introduced in the study of physical systems \cite{beck,wilkprl}. As shown in Ref.~\cite{tbssrpphysica}, superstatistics provides a derivation of the effective exponential model, that is dual to the Takahashi $q$-exponential model given by Eq.~\eqref{Takahashimodel}.

\begin{wrapfigure}{l}{0.4\textwidth}
  \vspace{-10pt}
   \centering
   {\includegraphics[width=0.38\textwidth]{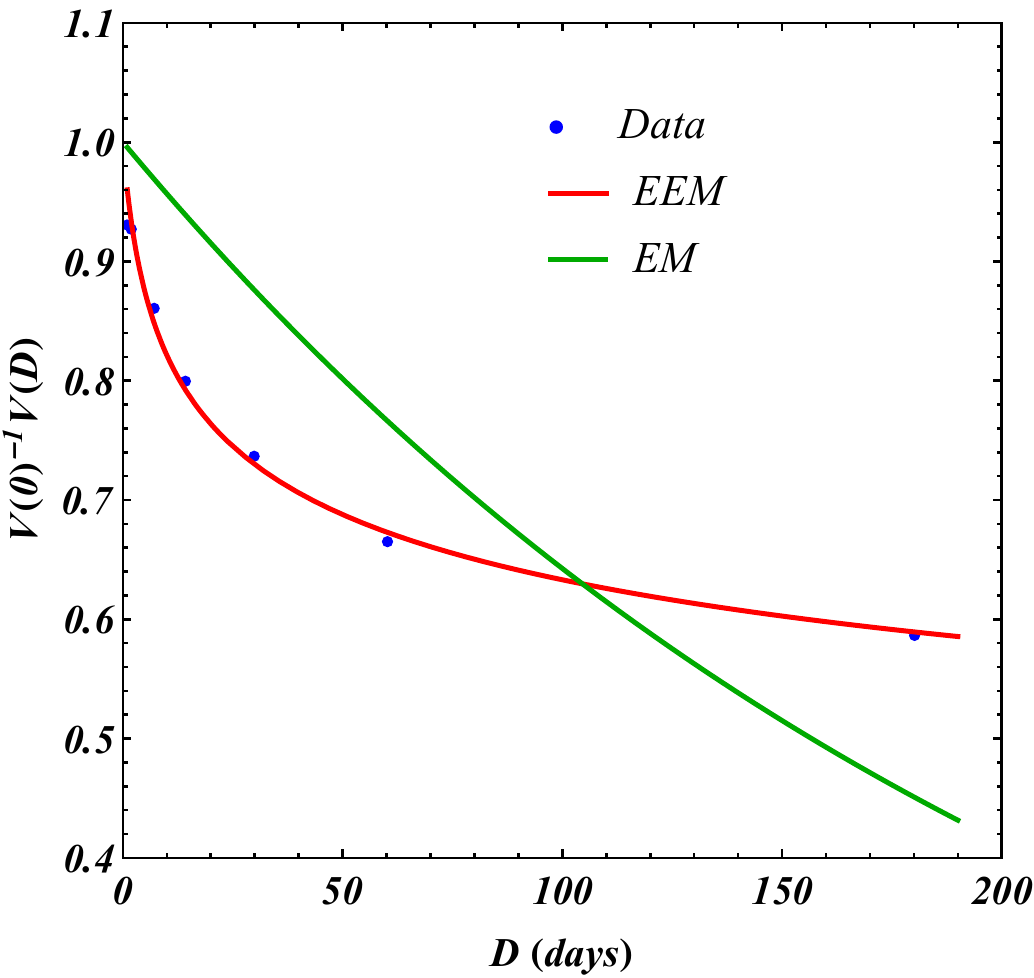}}
  \vspace{-10pt}
\caption{Fitting observed data obtained from Ref.~\cite{mansouridata} using the EM and EEM. EM: $\kappa_0 = 0.0044 \pm 0.0015$.
EEM: $q=9.067 \pm 0.116,~\kappa_1=0.048 \pm 0.013$.}
\label{mansourifiteem}
 \vspace{-10pt}
\end{wrapfigure}Superstatistics considers that impulsivity, which has long been used in various models of delay
discounting as a scale parameter, is a positive and fluctuating quantity. This comes as no surprise as within a social group impulsivity does vary. 
With this observation, the Effective Exponential Model (EEM), which contains a power-law function resembling the one discussed by Constantino Tsallis \cite{Tsal88}, was formulated in Ref.~\cite{tbssrpphysica}. It was shown that the $q$-exponential model of delay discounting proposed by Takahashi in Ref.~\cite{takahashi1} is dual to the effective exponential model in Ref.~\cite{tbssrpphysica} by a $q\leftrightarrow2-q$ transformation. Hence, the effective exponential and the Takahashi model are similar, but not exactly the same in appearance. Notwithstanding, Takahashi's model was not derived from superstatistics, and hence the approach followed in Ref.~\cite{tbssrpphysica} (and to be adapted in the present article) differs. 

At this point, it will be worthwhile to briefly recapitulate some key details of the EEM. It was motivated that in a social system, impulsivity as a positive random variable should have a distribution. In the absence of any more knowledge of how the impulsivity distribution may look like, we may choose one of the `least informative options' -- the gamma distribution \cite{hogg,tsqgaussian}. 

By weighting impulsivity $\kappa$ in the exponential model (EM) of delay discounting in Eq.~\eqref{expmodel} with the gamma distribution, the effective exponential factor, that looks like the Tsallis-like $q$-exponential power-law function, was obtained.  The EEM proposed in Ref.~\cite{tbssrpphysica} is given by
\bea
V(D) &=& V(0) \exp_q(-\kappa_1 D) \nn\\
&=& \frac{V(0)} {\left(1+(q-1)\kappa_1 D\right)^{\frac{1}{q-1}}}. ~~~~~~\text{(--EEM--)} 
\label{ourTsmodel}
\eea
In the above equation, $q$ may be called the `nonextensivity parameter' and $\kappa_1$ is impulsivity. In Ref.~\cite{tbssrpphysica}, they were shown to be related to the relative variance and the mean of impulsivity respectively. It is also straightforward to verify the $q\leftrightarrow2-q$ duality between Eqs.~\eqref{Takahashimodel} and \eqref{ourTsmodel}. 
Fig.~\eqref{mansourifiteem} shows a comparison of the EM and the EEM while describing experimental data.
\\
\\
{\it Statement of the problem}: This article aims to extend the Effective Exponential Model by incorporating impulsivity as a bi-faceted construct. One important question that may be asked in this context is how to combine the two facets to get a resultant value. For example, if the two facets of impulsivity are trait and state impulsivities, how do they combine? One popular choice is to add the two facets. We show that such an operation to combine the facets also lead to a model similar to the effective exponential model in Eq.~\eqref{ourTsmodel}. However, we also explore the possibility of other operations to combine two facets of impulsivity.
The resulting group of models may be referred to as the Extended Effective Exponential Models or E$^3$M. To the best of our knowledge, this is the first attempt to derive a mathematical model of delay discounting considering that impulsivity may be a multi-faceted quantity. Furthermore, the novelty of this paper lies in the utilization of the superstatistics approach that considers fluctuation in scale parameters (e. g. temperature in a plasma or impulsivity in a social system). Similar to the EEM, the E$^3$M identifies a quasi power-law factor that can represent temporal discounting. In the following section, we will delve into the details of the steps to obtain such a factor for both additive and nonadditive cases.

\section{Additive model}
In this model we consider impulsivity $\kappa$ to be described by an unweighted addition of two facets $\kt$ and $\ks$. Hence, the exponential factor in Eq.~\eqref{expmodel} will simply be
$\exp\left(-(\kt+\ks)D\right)$. $\kt$ and $\ks$ are fluctuating quantities such distributed according to a gamma distribution,
\bea
f(k_i) = \Gamma\left(\frac{1}{q_i-1}\right)
\left\{\frac{1}{(q_i-1)k_{i}^0}\right\}^{\frac{1}{q_i-1}-1} \exp\left(-\frac{k_i}{(q_i-1)k_{i}^0}\right)
(i: \text{T}, \text{S}).
\eea
Following Ref.~\cite{tbssrpphysica} we can define the effective exponential factor
\bea
\int_0^{\infty} dk_i ~f(k_i) \exp\left(-k_i D\right) = \left(1+(q_i-1)k_i^0 D\right)^{-\frac{1}{q_i-1}}
\eea
that is nothing but a $q$-exponential function. Hence the additive model would be given by the following equation: 
\bea
V(D) = V(0) \left(1+(\qt-1)\kt^0 D\right)^{-\frac{1}{\qt-1}} \left(1+(\qs-1)\ks^0 D\right)^{-\frac{1}{\qs-1}}.
\eea
However, it is possible to show that addition is not the only operator that can combine two facets of impulsivity. In the next section we explore this aspect.


\section{Nonadditive model}

\label{model}

\subsection{Step 1: finding a joint distribution}
Following the previous section, we consider impulsivity to be represented by $\kt$ and $\ks$, two positive, independent random variables. In the absence of any further knowledge of how they are distributed in a (social) system, one of the least informative options, {\it i.e.}, a joint gamma distribution of $\kt$ and $\ks$ given by \cite{hogg,tsqgaussian}
\bea
f(\kt,\ks) =  \frac {\kt^{\frac{1}{\qt-1}-1} \ks^{\frac{1}{\qs-1}-1} \exp(-\frac{\kt+\ks}{\theta})} {\theta^{\frac{1}{\qt-1}} \theta^{\frac{1}{\qs-1}} 
\Gamma\left(\frac{1}{\qt-1} \right) 
\Gamma\left(\frac{1}{\qs-1}\right)},
\label{jointgamma}
\eea
for a single scale parameter $\theta>0$ and shape parameters $1<\qt<\infty$ and $1<\qs<\infty$, may be considered.

Now, we consider that impulsivity $\kr$ is formed by a (nonadditive) combination of $\kt$ and $\ks$. Hence, we have to find out the distribution of the new positive, random variable $\kr$. Such a ritual is clearly described in textbooks of statistics \cite{hogg} and involves introducing an auxiliary random variable $\ka = \kt + \ks$. So, the variable transformation we are looking for is $\{\kt,\ks\}\rightarrow\{\kr,\ka\}$. The distribution in terms of $\kr$ and $\ka$ is related to that of $\kt$ and $\ks$ (Eq.~\ref{jointgamma}) by the following relation:
\bea
f(\kr,\ka) = f[\kt(\kr,\ka),\ks(\kr,\ka)] \left| \text{det}~J\right|,
\label{fkrka1}
\eea
where $\left| \text{det}~J\right|$ is the absolute value of the determinant of the Jacobian variable transformation matrix defined by
\bea
J = \begin{pmatrix}
\frac{\partial \ks} {\partial \kr} & \frac{\partial \ks} {\partial \ka} 
\\
\\
\frac{\partial \kt} {\partial \kr} & \frac{\partial \kt} {\partial \ka}
\end{pmatrix}
\label{jacobian}
\eea
It is noteworthy that by choosing the auxiliary variable to be $\ka = \kt + \ks$, we exclude the possibility that $\kr = \kt + \ks$, as in that case the Jacobian would be zero.
It is apparent from Eq.~\eqref{jacobian} that to find the joint distribution we must know what $\kr$ is. So far, we have just mentioned that $\kr$ is formed
out of a nonadditive combination of $\kt$ and $\ks$, but did not explore what exactly that combination may be. We discuss this question in the next section.

\subsection{Step 2: combining $\kt$ and $\ks$}
How the two facets of impulsivity combine does not have a straightforward answer. Although to provide utility they are added in general, the real scenario may be different. 
To begin with, we introduce examples used in Ref.~\cite{tsqgaussian} that considers two impulsed moments induced by fluctuations in a medium. In terms of financial markets two random variables may be similar to predicted increase or decrease of stock prices. So, fluctuation in price is given by the difference of the two variables. Such a system's response will be dependent on these variables and can be represented by some kind of mean ($\mathcal{M}$) between the two variables. 
So following the analogy of Ref.~\cite{tsqgaussian}, one option is to treat $\kr$ as the following:
\bea
\kr = \mathcal{M}^{-1} (\kt,\ks) (\kt - \ks).
\label{krmean}
\eea
But adapting such an ansatz for $\kr$ in our problem leads to a few issues. The first issue lies in an absence of a direct analogy between two facets of impulsivity and increase/decrease of stock prices. Even if we assume the analogy to be valid, $\kr$ can be negative in Eq.~\eqref{krmean} whereas, according to our consideration, impulsivity is a (a) positive and (b) fluctuating variable ($0<\kr<\infty$). So, although the ansatz in Eq.~\eqref{krmean} satisfies condition (b), condition (a) is not satisfied. Ref.~\cite{tsqgaussian} also explores the combination $\kr = \kt (\kt+\ks)^{-1}$ that may be obtained from Eq.~\eqref{krmean}. But this ansatz for $\kr$, although yields a positive fluctuating variable, makes it bounded between 0 and 1. So, $\kr = \kt (\kt+\ks)^{-1}$ may be a possible candidate modulo the boundedness restriction. Further exploration brings us to the last option suggested in Ref.~\cite{tsqgaussian} given by $\kr = \kt \ks^{-1}$. As one notices, the last ansatz yields a positive, random variable that is not bounded, and it fulfils our criteria. Hence, in our analysis we choose
\bea
\kr = \frac{\kt}{\ks}.
\eea 

\subsection{Step 3: finding a joint distribution in terms of $\kr$}
Now, with $\kr = \kt \ks^{-1}$ and $\ka = \kt + \ks$, we find that
\bea
\kt &=& \ka \kr (\kr +1)^{-1};~  \ks = \ka(\kr+1)^{-1} ; \nn\\
\frac{\partial \ks} {\partial \kr} &=& -\ka (\kr+1)^{-2};~ \frac{\partial \ks} {\partial \ka} = (\kr+1)^{-1};~ \nn\\
\frac{\partial \kt} {\partial \kr} &=& \ka (\kr+1)^{-1} - \ka \kr (\kr+1)^{-2};~\nn\\
\frac{\partial \kt} {\partial \ka} &=& \kr (\kr+1)^{-1};~ \nn\\
\text{and~hence,}~ \left| \text{det}~J\right| &=& \ka (\kr+1)^{-2}.
\label{ktks}
\eea
Putting Eqs.~\eqref{ktks} in Eq.~\eqref{fkrka1}, we obtain the following joint distribution:

\bea
f(\kr,\ka) &=&  
\frac{\ka} {(\kr +1)^{2}}
\left( \frac{\ka \kr}{\kr +1} \right)^{\alpht-1}
\left( \frac{\ka}{\kr +1} \right)^{\alphs-1} 
\frac {\exp\left( -\frac{\ka}{\theta} \right)} {\theta^{\alpht + \alphs} \Gamma\left(\alpht\right) \Gamma\left(\alphs\right) }\nn\\
&=& \frac{ \kr^{\alpht-1}  \ka^{\alpht+\alphs-1} }   {(\kr +1)^{\alpht+\alphs}}  
\frac {\exp\left( -\frac{\ka}{\theta} \right)} {\theta^{\alpht + \alphs} \Gamma\left(\alpht\right) \Gamma\left(\alphs\right) },
\label{fkrka2}
\eea
where we have defined two positive quantities $\alphs = (\qs -1)^{-1}$ and $\alpht = (\qt-1)^{-1}$.

\subsection{Step 4: Finding the extended effective exponential factor}
We noticed earlier \cite{tbssrpphysica} that if we replace the exponential factor in the exponential model in Eq.~\eqref{expmodel} with a weighted average, we get the quasi power-law
effective exponential model given by Eq.~\eqref{ourTsmodel}. Similarly, to obtain a nonadditive (i.e. the facets are not added) model we define an `extended effective exponential factor' (essentially an average), for a scaled variable $D$,
defined by
\bea
\epsilon_{\text{eff}}^{\text{ext}} 
&=& \frac {\int_0^{\infty} d\kr \exp(-\kr D) f(\kr,\ka)} {\int_0^{\infty} d\kr f(\kr,\ka)} \nn\\
&=& \frac {\int_0^{\infty} d\kr \exp(-\kr D) \kr^{\alpht-1} (\kr +1)^{-\alpht - \alphs} }
{\int_0^{\infty} d\kr \kr^{\alpht-1} (\kr +1)^{-\alpht - \alphs}} 
\label{fact1} \\
&=& \frac{\Gamma(\alphs + \alpht) }{\Gamma(\alphs)} ~\mathcal{U} (\alpht,1-\alphs,D),
\label{fact2}
\eea
and replace the exponential factor of the exponential model with $\epsilon_{\text{eff}}^{\text{ext}}$. Hence the E$^3$M can be written as
\bea
V(D) 
&=& V(0) \times \epsilon_{\text{eff}}^{\text{ext}} = V(0)  ~\mathcal{U} (\alpht,1-\alphs,D) ~\frac{\Gamma(\alphs + \alpht) }{\Gamma(\alphs)} \nn\\
&=& V(0)  ~\mathcal{U} \left( \frac{1}{\qt-1},\frac{\qs-2}{\qs-1},D \right) ~\frac{\Gamma \left( \frac{1}{\qs-1} + \frac{1}{\qt-1} \right) } { \Gamma\left(\frac{1}{\qs-1}\right) }.
~~~~~~\text{(--E$^3$M--)}
\label{e3m}
\eea
In the above equation, $ \Gamma$ is the gamma function represented by the integral \cite{Abramgamma}
\bea
\Gamma(n) = \int_0^{\infty} dx \exp(-x) x^{n-1};~\forall~\mathcal{R}(n)>0,
\eea
and $\mathcal{U}$ is the confluent hypergeometric function of the second kind represented by the integral \cite{Abramconfhyp}
\bea
\mathcal{U}(a,b,z) = \frac{1}{\Gamma(a)} \int_0^{\infty} dt \exp(-z t) t^{a-1} (1+t)^{b-a-1};~\forall ~\mathcal{R}(a),\mathcal{R}(z)>0.
\eea
Hence, the numerator of Eq.~\eqref{fact1} yields the confluent hypergeometric function $\mathcal{U}$. On the other hand,
the denominator of Eq.~\eqref{fact1} can be calculated in terms of the Beta function represented by \cite{Abrambeta}
\bea
\mathcal{B} (m+1, n+1) = \int_0^{\infty} \frac {\hspace{-20pt} u^m du}{(1+u)^{m+n+2}} = \frac{\Gamma(m+1)\Gamma(n+1)}{\Gamma(m+n+2)}.
\label{betafunc}
\eea
Eq.~\eqref{e3m} is the main result of our paper. 

\begin{figure}[!htb]
\minipage{0.45\textwidth}
\includegraphics[width=\linewidth]{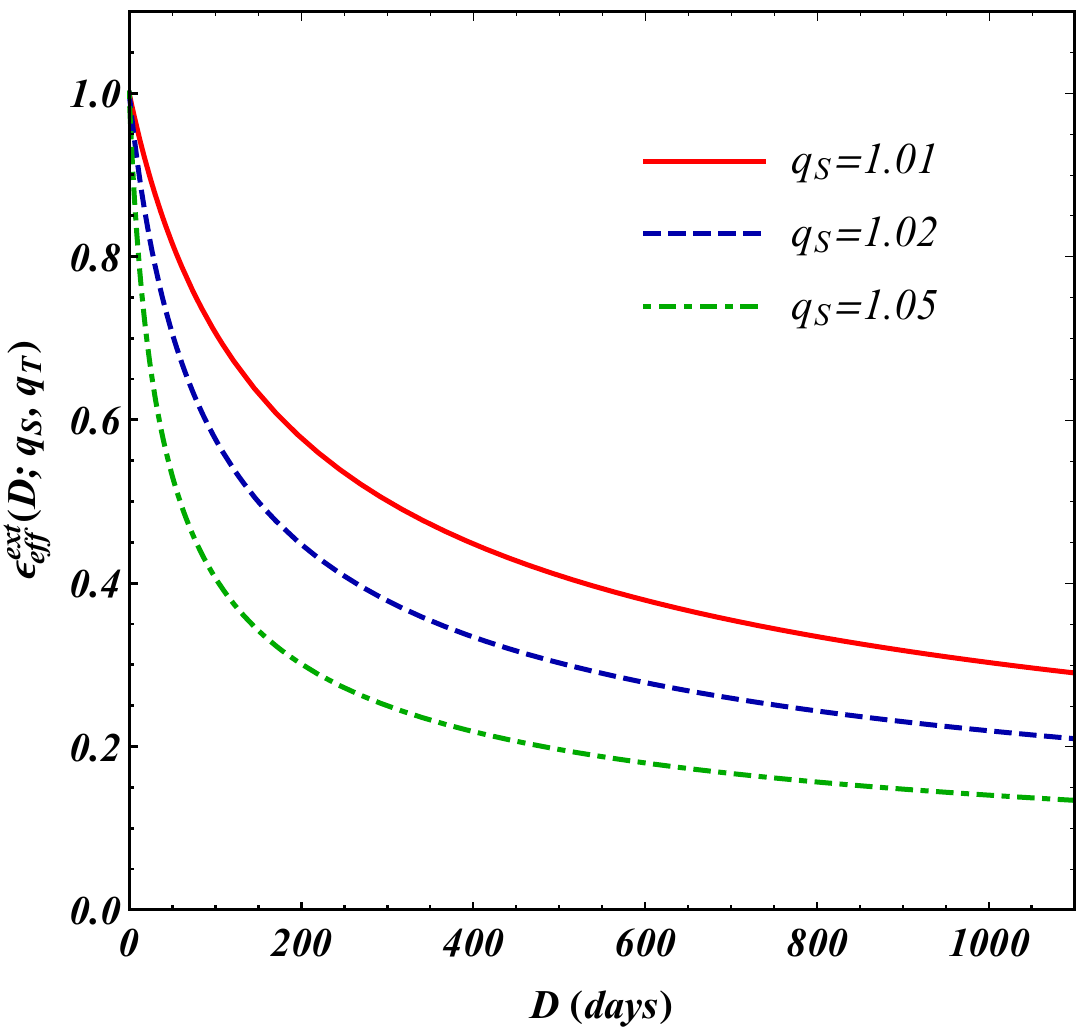}
\caption{Variation $\epsilon_{\text{eff}}^{\text{ext}}$ with $D$ for different $\qs$ values and for a fixed $\qt=3.01$.}
\label{epseffqs}
\endminipage\hfill
\minipage{0.45\textwidth}
\hspace{-0.1cm}
\includegraphics[width=\linewidth]{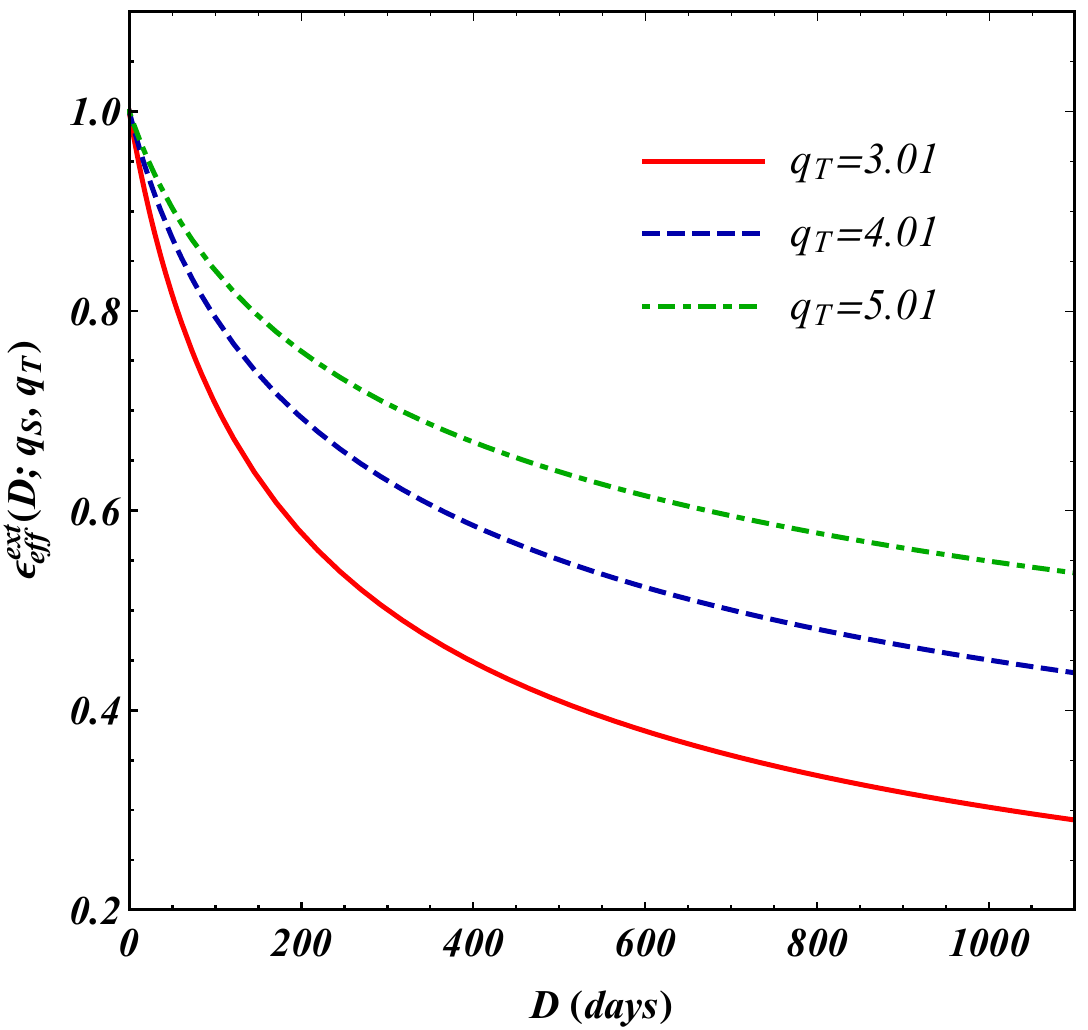}
\caption{Variation $\epsilon_{\text{eff}}^{\text{ext}}$ with $D$ for different $\qt$ values and for a fixed $\qs=1.01$.}
\label{epseffqt}
\endminipage\hfill
\end{figure}

To further explore the result, variation of the factor $\epsilon_{\text{eff}}^{\text{ext}}$ with delay is shown in Figs.~\eqref{epseffqs} and \eqref{epseffqt}. We observe that the factor monotonically decreases with delay and hence is suitable to describe the gradual decrease of indifference points in experimental observables. The average value of the variable $\kr$ is given by
\bea
\langle \kr \rangle =  \left\langle \frac{\kt}{\ks} \right\rangle &=& \frac {\int_0^{\infty} d\kr \kr f(\kr,\ka)} {\int_0^{\infty} d\kr f(\kr,\ka)} \nn\\
&=& \frac {\int_0^{\infty} d\kr \kr^{\alpht} (\kr +1)^{-\alpht - \alphs} }
{\int_0^{\infty} d\kr \kr^{\alpht-1} (\kr +1)^{-\alpht - \alphs}} \nn\\
&=& \frac{\alpht}{\alphs-1} = \frac{\qs-1}{(\qt-1)(2-\qs)} ~~~~~\text{(using Eq.~\ref{betafunc})}.
\label{krav}
\eea
Since $\kr$ is a positive variable, its average should be positive. This imposes a new constraint $\qs<2$ for Eq.~\eqref{krav} to be satisfied (since $\qs,\qt>1$).
In section \ref{data}, we will consider a set of observed data to investigate the applicability of the model. 

\section{Data analysis using the E$^3$M}
\label{data}

{\it Methodology}: We utilize our model in Eq.~\eqref{e3m} to explain existing observed data for indifference points in delay discounting tasks. 
We consider eight open-access datasets from three different studies (\textcolor{black}{two longitudinal and one cross-sectional}).  Datasets are analyzed with the help of C++ codes utilizing the MINUIT package of the ROOT program \cite{root}. For visualization purposes and additional verification of results, the Mathematica software \cite{mathematica} has been used. 

\begin{figure}[h]
\minipage{0.45\textwidth}
\includegraphics[width=\linewidth]{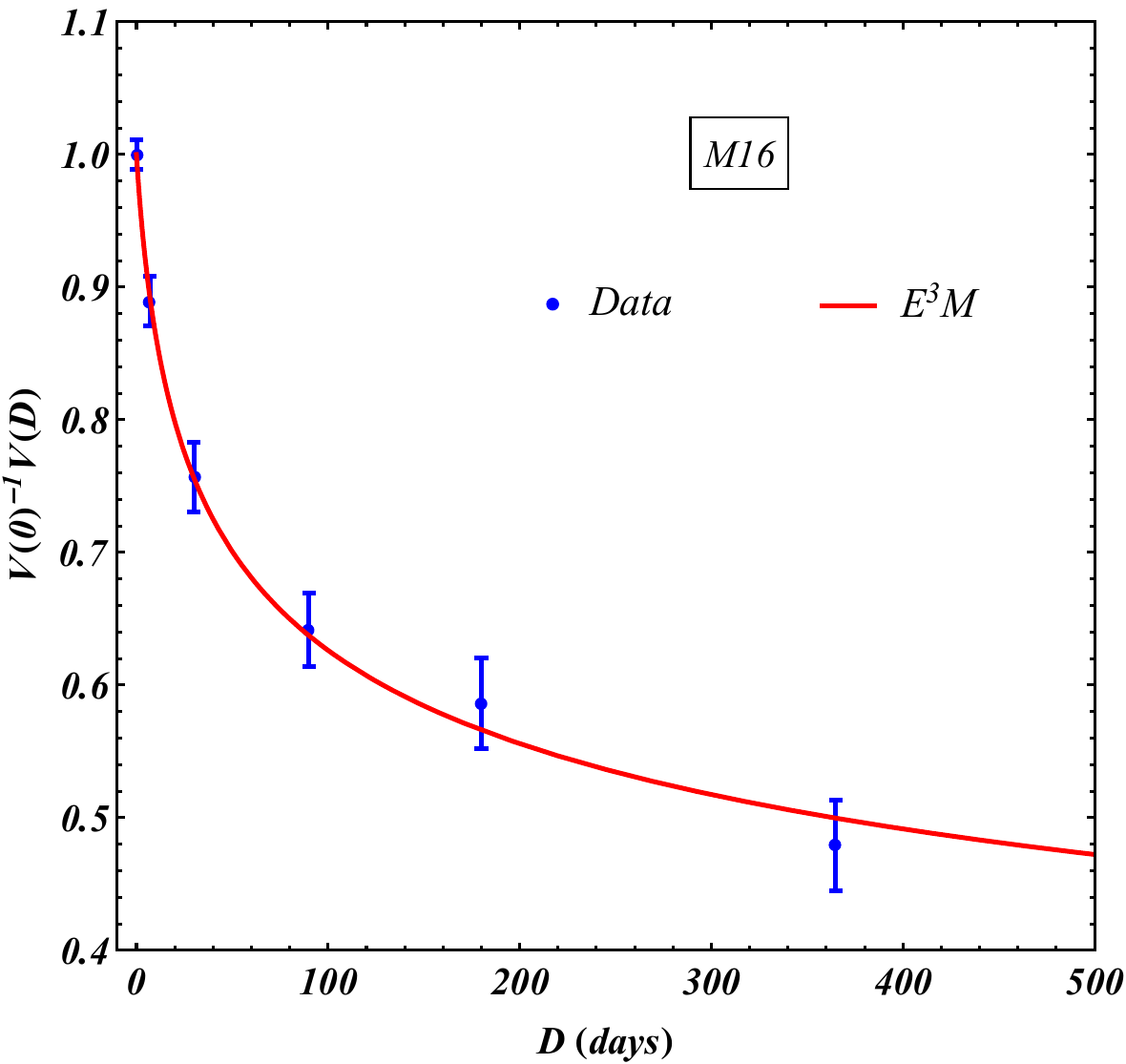}
\caption{Fitting longitudinal data obtained from Ref.~\cite{anokhin} for monozygotic twins at age 16 using the E$^3$M: $\qs=1.113 \pm 0.040,~\qt= 6.463\pm 0.816$.}
\label{mz16}
\endminipage\hfill
\minipage{0.45\textwidth}
\hspace{-0.1cm}
\includegraphics[width=\linewidth]{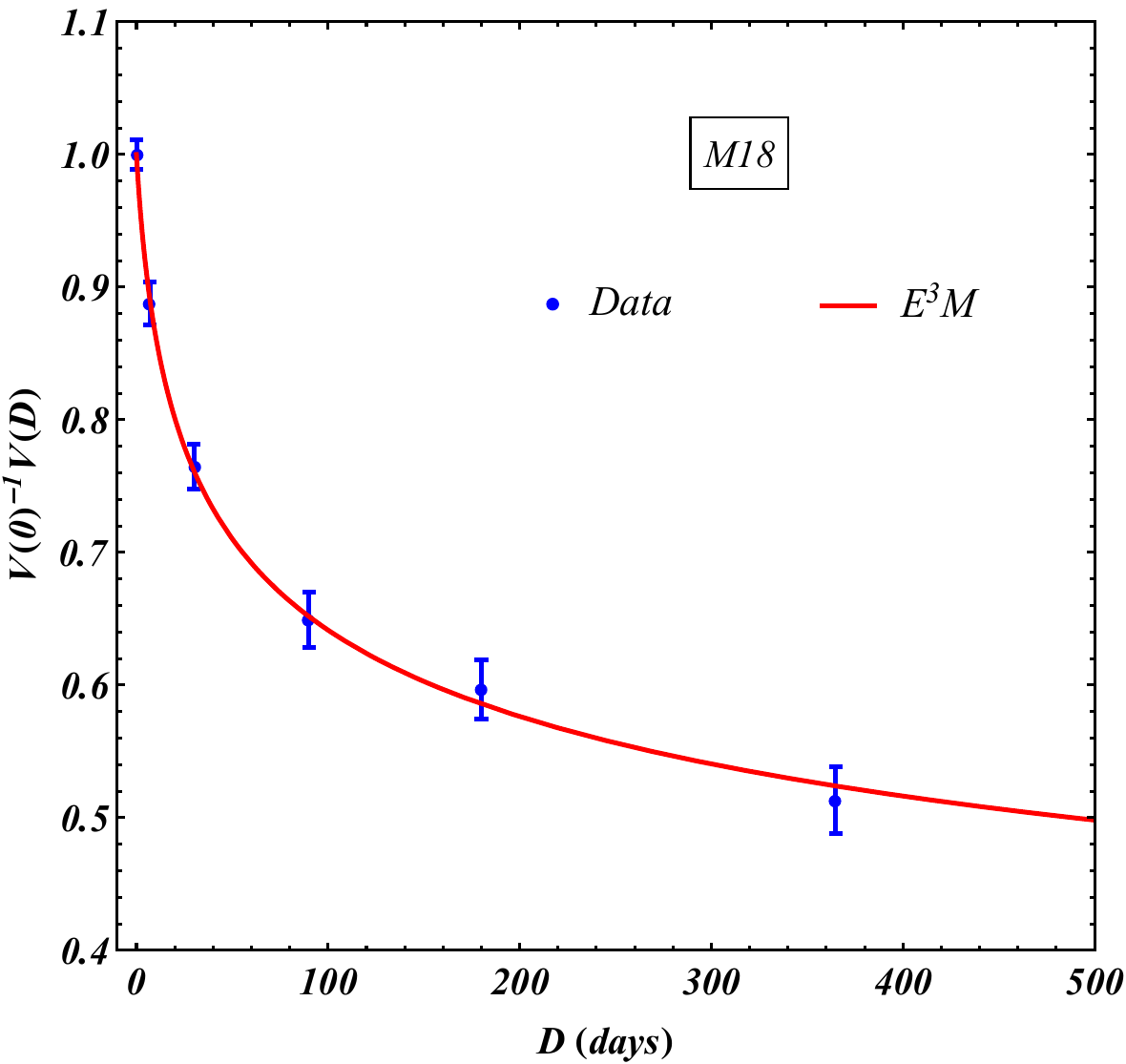}
\caption{Fitting longitudinal data obtained from Ref.~\cite{anokhin} for monozygotic twins at age 18 using the E$^3$M: $\qs= 1.134\pm 0.038,~\qt= 7.137\pm 0.705$.}
\label{mz18}
\endminipage\hfill
\minipage{0.45\textwidth}
\includegraphics[width=\linewidth]{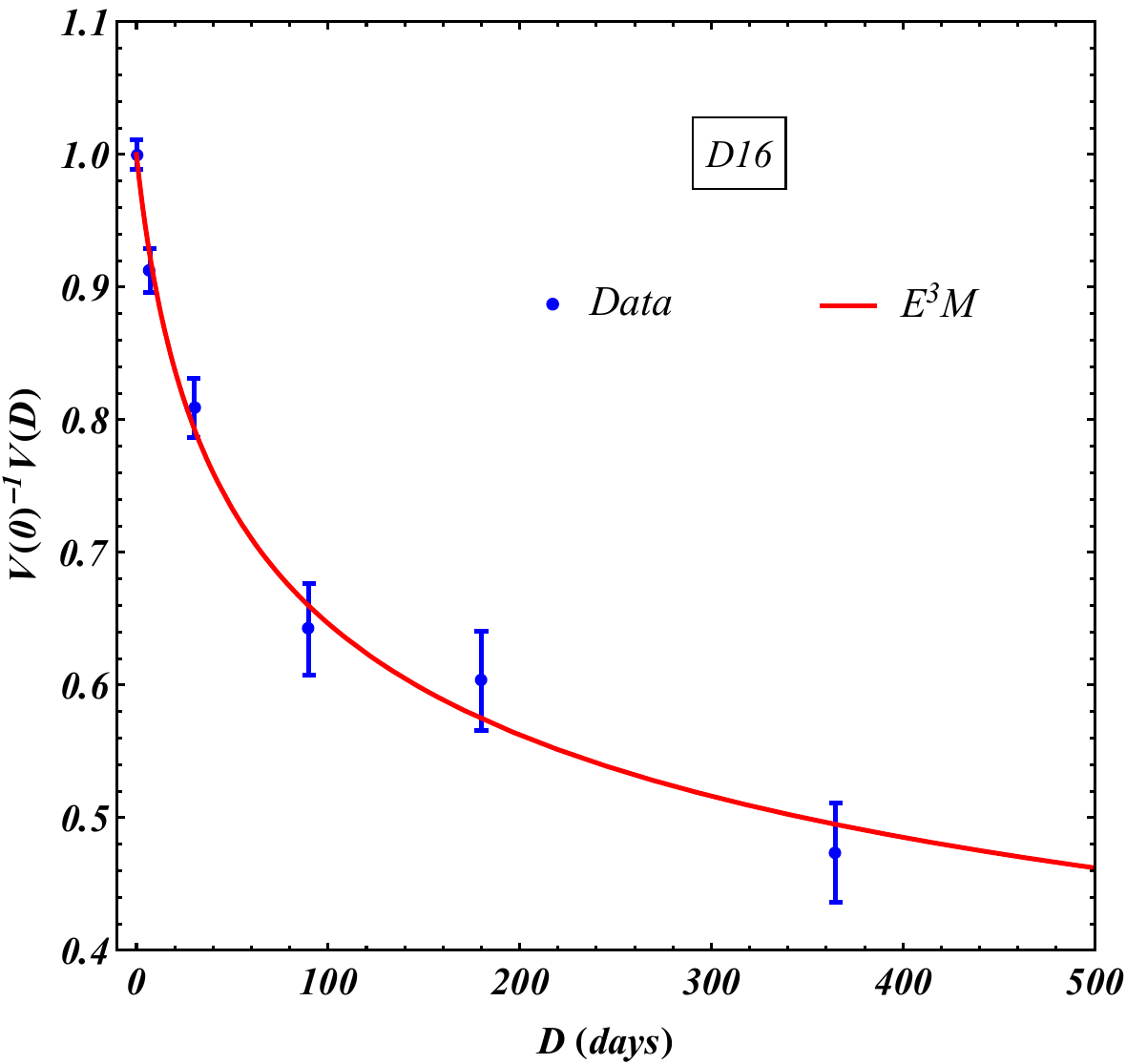}
\caption{Fitting longitudinal data obtained from Ref.~\cite{anokhin} for dizygotic twins at age 16 using the E$^3$M: $\qs= 1.057 \pm 0.021,~\qt= 5.414\pm 0.800$.}
\label{dz16}
\endminipage\hfill
\minipage{0.45\textwidth}
\hspace{-0.1cm}
\includegraphics[width=\linewidth]{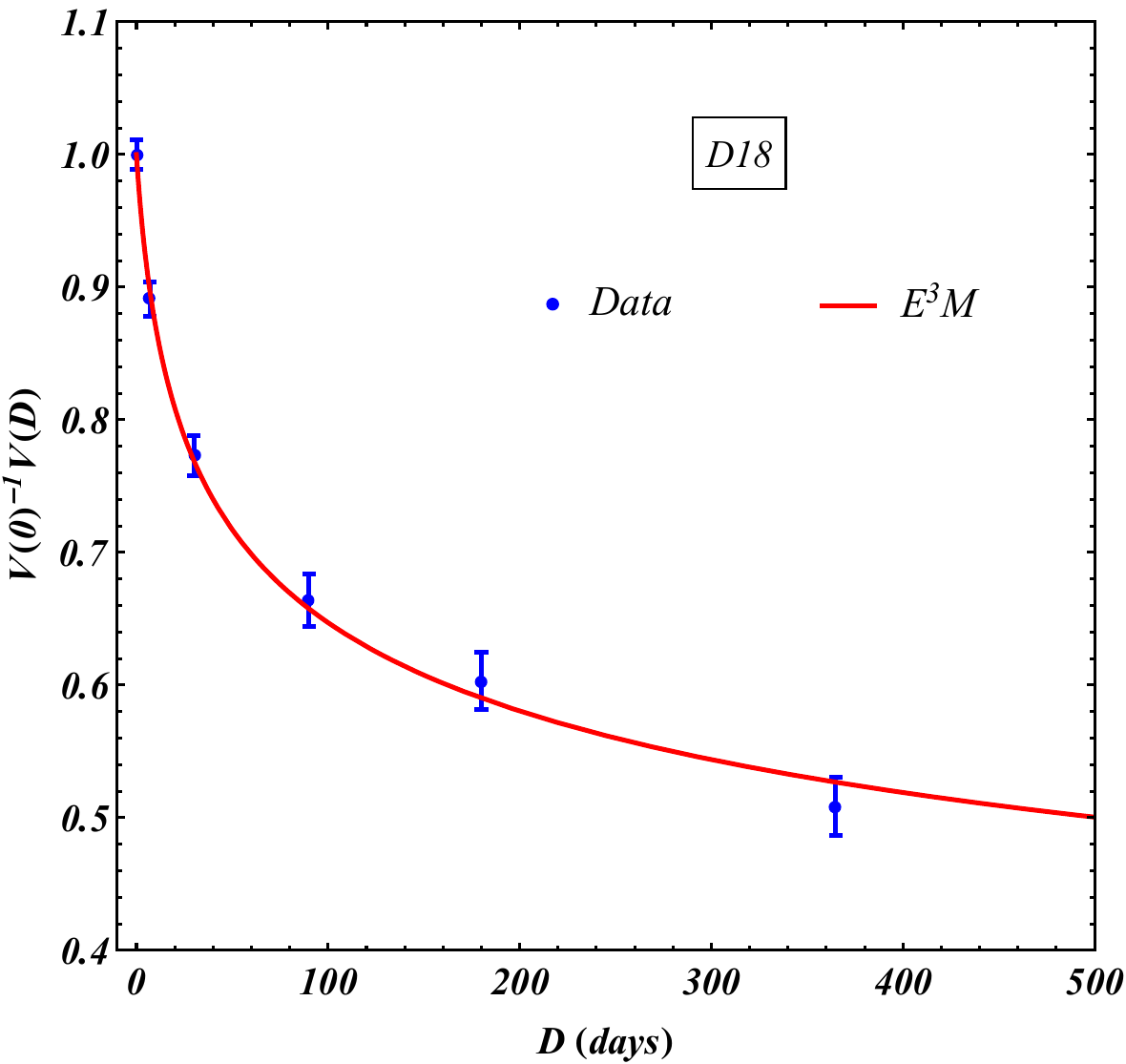}
\caption{Fitting longitudinal data obtained from Ref.~\cite{anokhin} for dizygotic twins at age 18 using the E$^3$M: $\qs= 1.120\pm0.030, ~\qt= 7.011\pm 0.624$.}
\label{dz18}
\endminipage\hfill
\end{figure}

\begin{figure}[h]
\minipage{0.45\textwidth}
\includegraphics[width=\linewidth]{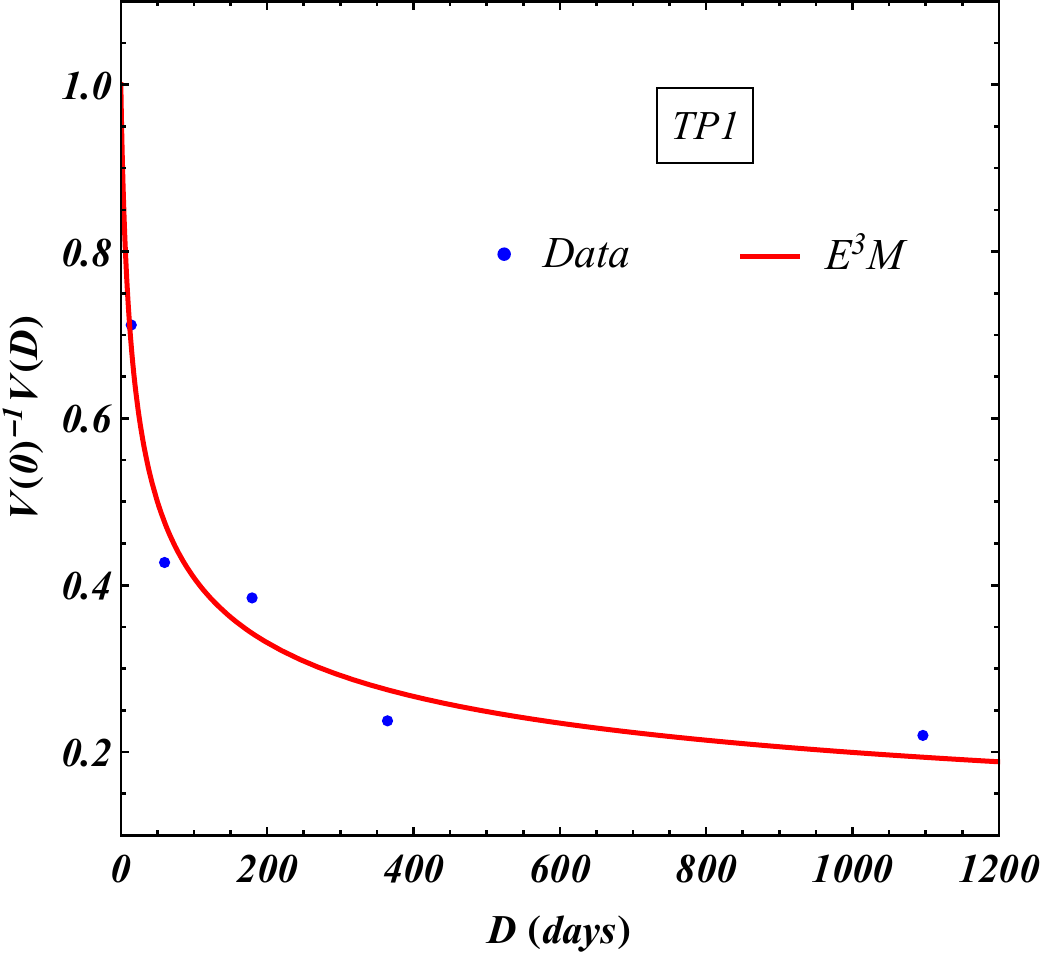}
\caption{Fitting longitudinal data obtained from Ref.~\cite{ordata} at time point 1 using the E$^3$M: $\qs=1.1445 \pm 0.068,~\qt=4.121 \pm 0.286$.}
\label{time1}
\endminipage\hfill
\minipage{0.45\textwidth}
\includegraphics[width=\linewidth]{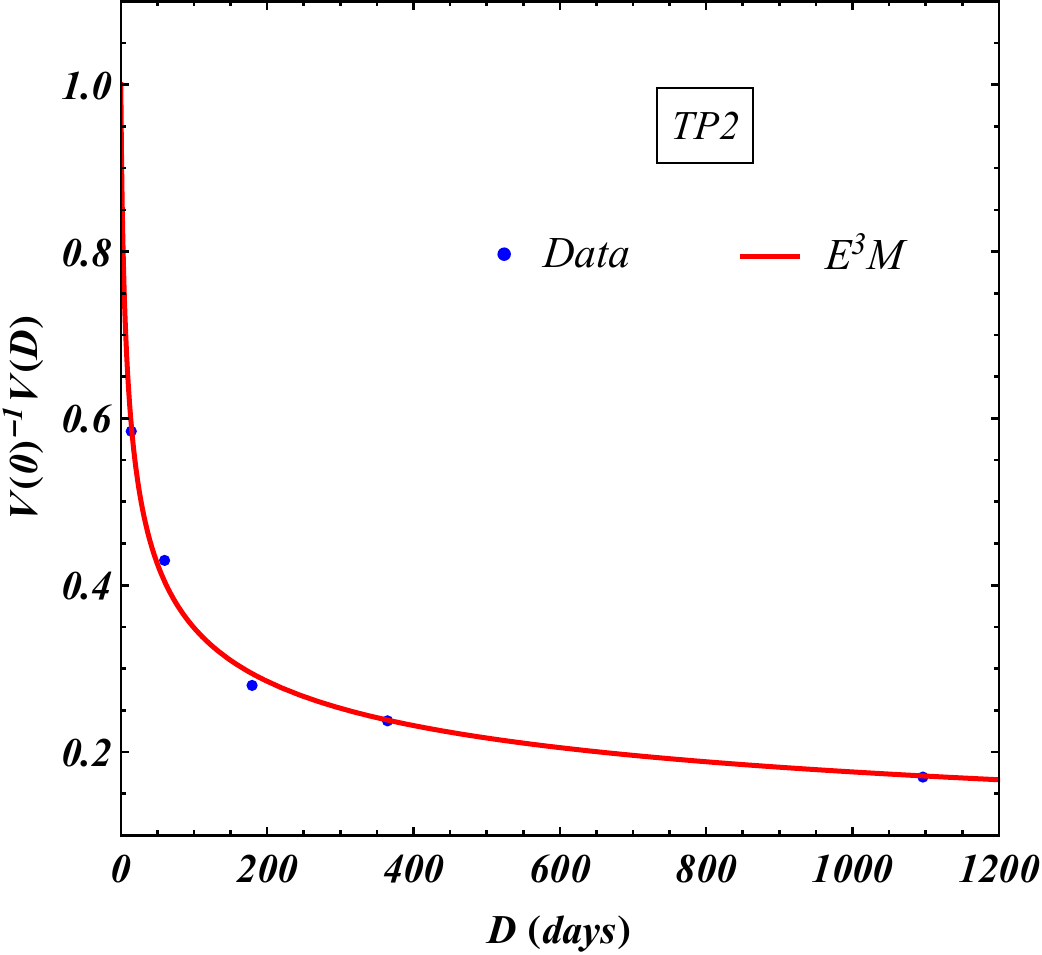}
\caption{Fitting longitudinal data obtained from Ref.~\cite{ordata} at time point 2 using the E$^3$M: $\qs=1.287 \pm 0.050,~\qt=4.320 \pm 0.100$.}
\label{time2}
\endminipage\hfill
\minipage{0.45\textwidth}
\vspace{20pt}
\includegraphics[width=\linewidth]{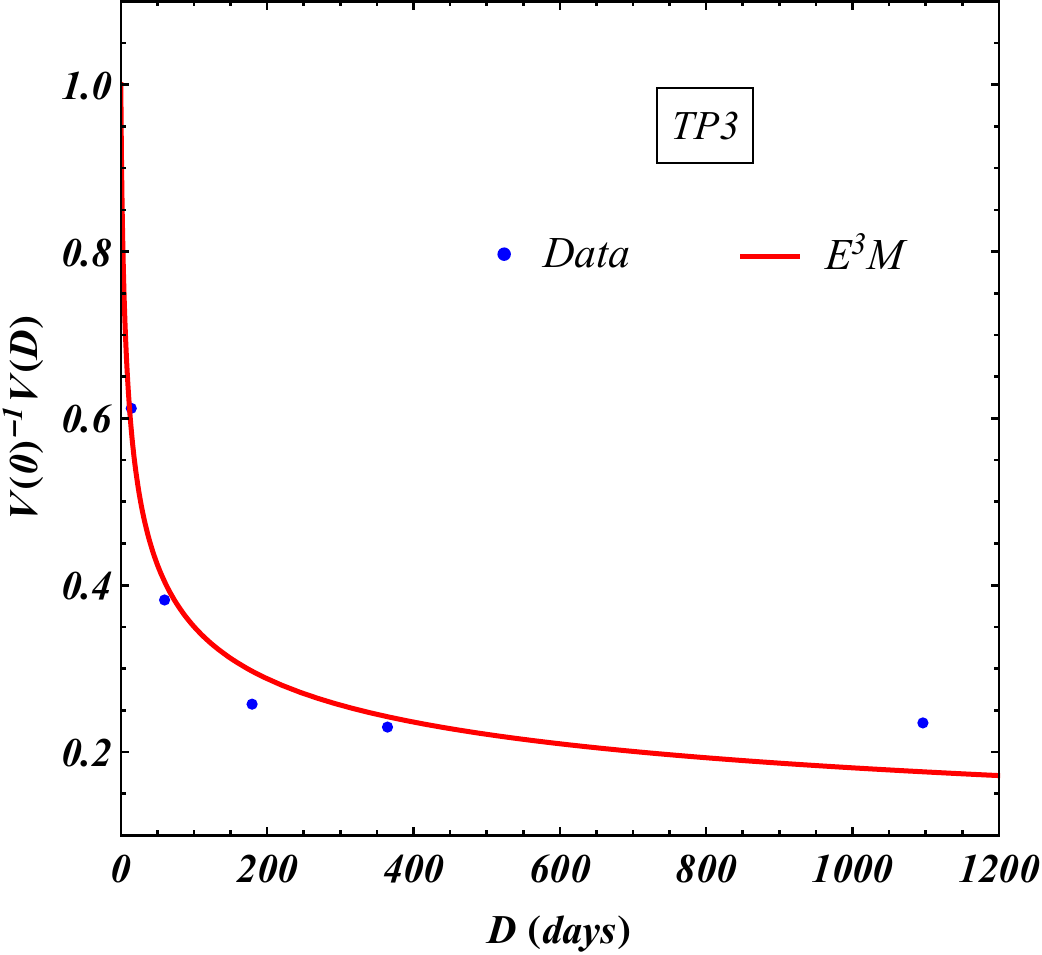}
\caption{Fitting longitudinal data obtained from Ref.~\cite{ordata} at time point 3 using the E$^3$M: $\qs=1.321 \pm 0.118,~\qt=4.453 \pm 0.107$.}
\label{time3}
\endminipage\hfill
\end{figure}

\begin{figure}[h]
  \vspace{-0pt}
   \centering
   {\includegraphics[width=0.45\textwidth]{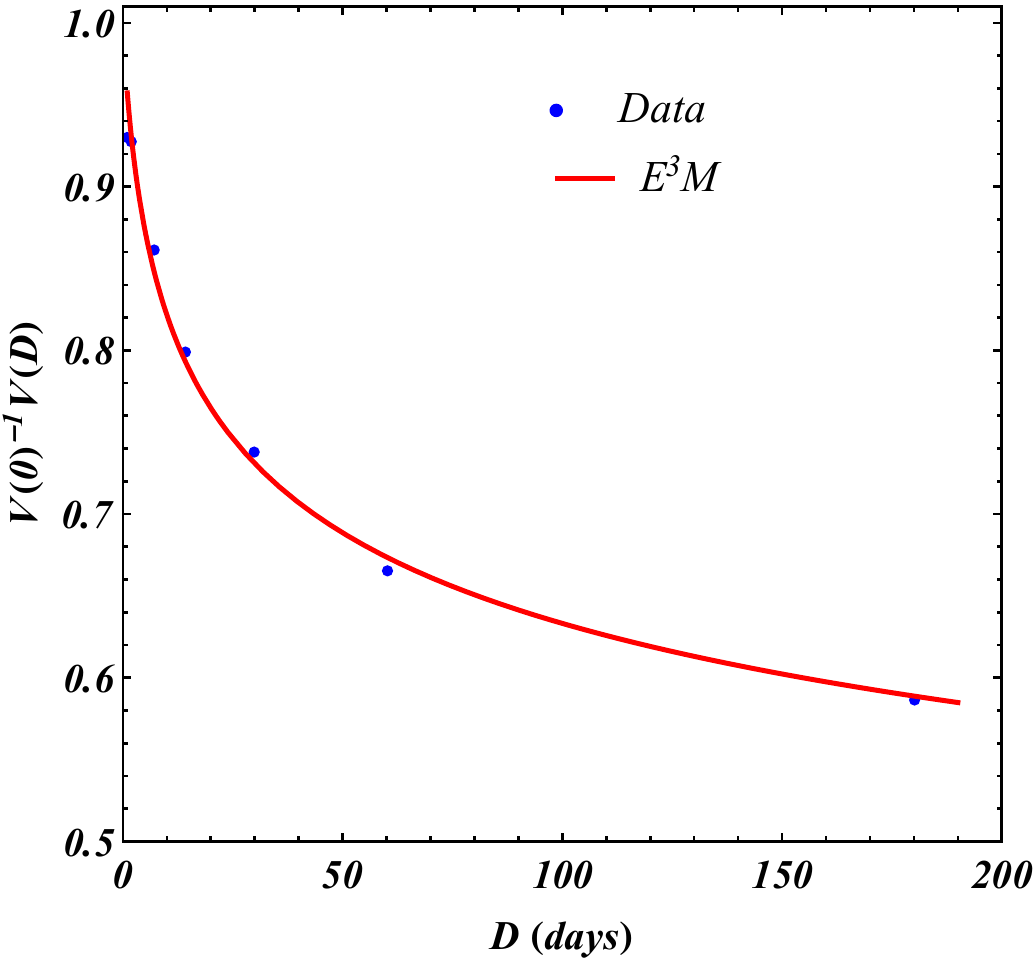}}
\vspace{-10pt}
\caption{Fitting observed data obtained from Ref.~\cite{mansouridata} using the E$^3$M: $\qs=1.310 \pm 0.076,~\qt=8.893 \pm 0.397$.}
\label{mansourifite3m}
 \vspace{-0pt}
\end{figure}

{\it Findings}: Considering the longitudinal data set associated with Refs.~\cite{anokhin} (Study 1) and \cite{ordata} (Study 2), we obtain Figs.~\eqref{mz16}-\eqref{time3}. Additionally, we utilize the cross-sectional data set associated with Ref.~\cite{mansouridata} (Study 3) and obtain Fig.~\eqref{mansourifite3m}. The indifference points obtained from the experiments have been averaged over all the participants. Details of the studies yielding the datasets utilized in this paper are provided in Table~\ref{tabdesstat}. We observe that our
model closely follows the data points in all these plots. Table~\ref{tabsum} summarizes the results obtained from the description of the datasets using the E$^3$M.

\begin{figure}[h]
\minipage{0.45\textwidth}
\includegraphics[width=\linewidth]{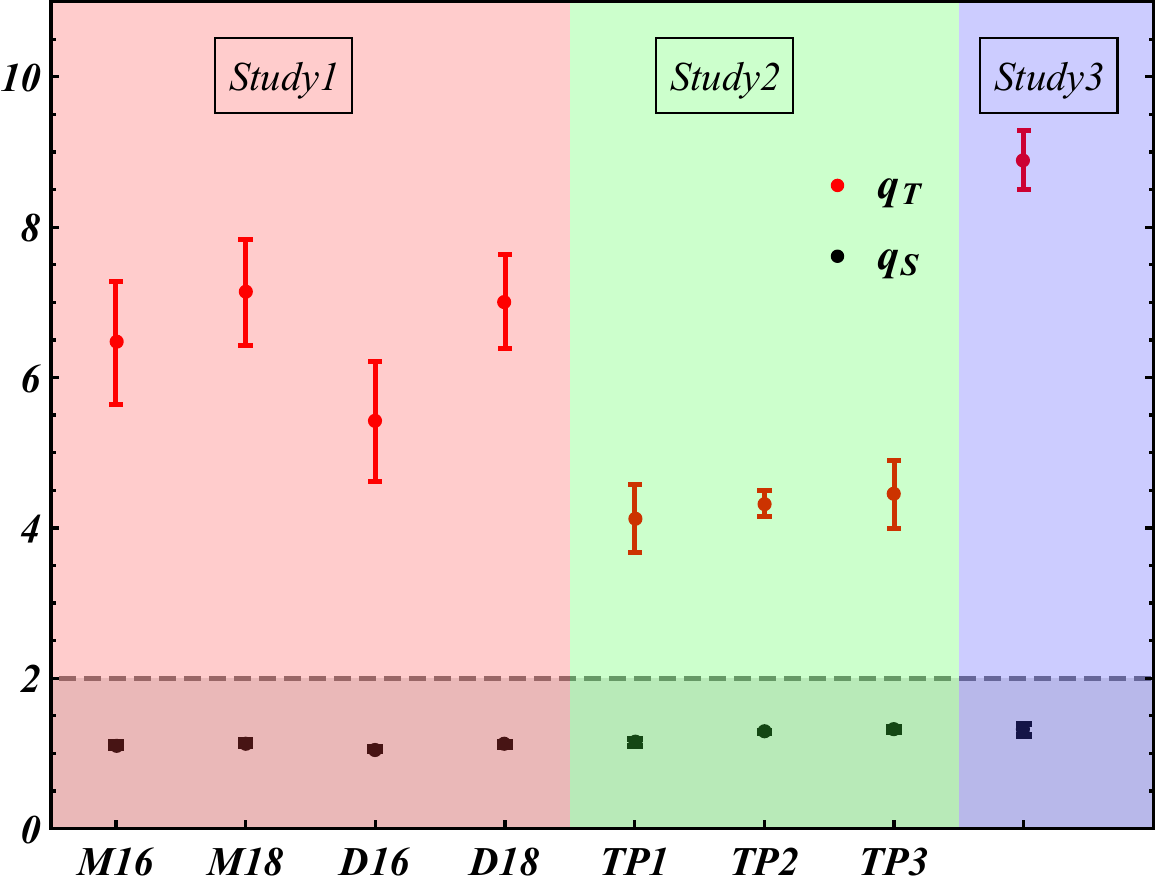}
\caption{Values of $\qs$ and $\qt$ obtained by fitting observed data utilizing the E$^3$M.}
\label{qsqtall}
\endminipage\hfill
\minipage{0.48\textwidth}
\vspace{-2pt}
\hspace{-0.21cm}
\includegraphics[width=\linewidth]{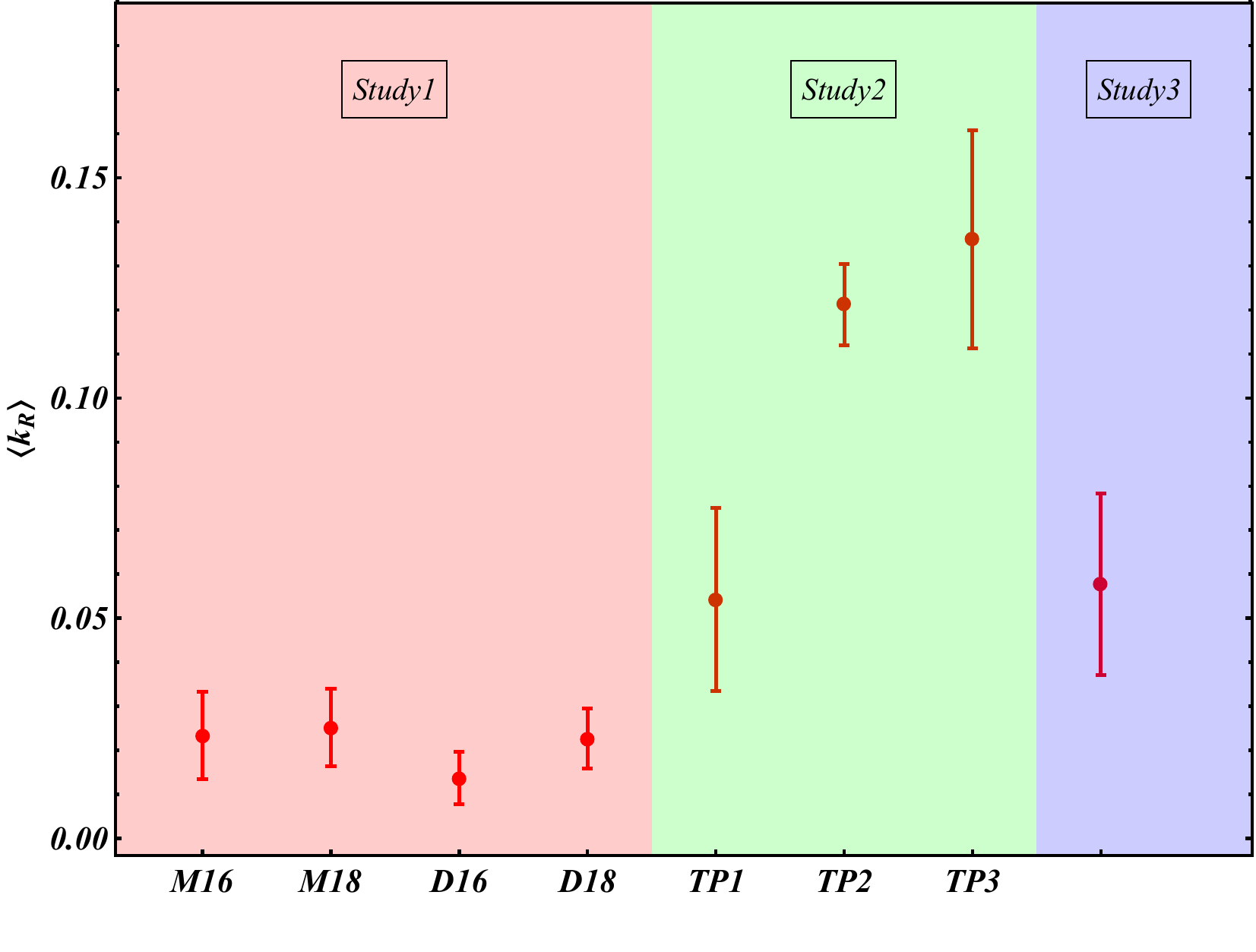}
\vspace{-0pt}
\caption{Values of $\langle \kr \rangle$ calculated from different studies.}
\label{rinhall}
\endminipage\hfill
\end{figure}

Figs.~\eqref{qsqtall} and \eqref{rinhall} graphically represent the parameter values obtained from the studies and $\langle \kr \rangle$. In these figures, Studies 1, 2, and 3 have been depicted by different shaded regions of red, green, and blue respectively. Interestingly, for these datasets, the additional constraint $\qs<2$, obtained from Eq.~\eqref{krav}, is followed and $\langle \kr \rangle<1$. 

From Fig.~\eqref{qsqtall}, it seems that the $\qs$ values (with error bars almost coinciding with the diameter of the data points used to represent the central values) are almost constant, but on closer inspection, we notice that the trend followed by $\qs$ (central values) is similar to that of $\qt$. Hence, $\langle \kr \rangle$ calculated from these parameter values also follow the same trend, as shown in Fig.~\eqref{rinhall}. In Study 1, the central value of $\langle \kr \rangle$ shows a dip for dizygotic twins at age 16 (D16). However, considering the error bars, a constant value of average impulsivity $\langle \kr \rangle$ can not be ruled out. On the other hand, for Study 2, the $\langle \kr \rangle$ value at time point 1 is distinctly separated from the other two, considering even the uncertainties. So, one observes that $\langle \kr \rangle$ monotonically increases. 

As $\langle \kr \rangle$ signifies the average ratio of two facets of impulsivity, one may be inclined to investigate which one of the facets dominate. One such example of two facets are trait and state impulsivity. However, in our analysis, identifying one of the facets as state impulsivity (and another as trait or vice versa) requires validation, and hence, from this study we can not comment on this issue.

We also observe that the $k_1$ parameter values obtained from the fitting of the same set of data (using Eq.~\eqref{ourTsmodel}) is (0.048 $\pm$ 0.013) and the $\langle \kr \rangle$ value is (0.058 $\pm$ 0.020). The $k_1$ parameter can be explained as the average impulsivity of the sample \cite{tbssrpphysica}, while $\langle \kr \rangle$ represents an average ratio.

\hspace{-10cm}\section{Summary, conclusion and outlook}
\label{summary}

In this article, we propose an extension of the effective exponential model forwarded in Ref.~\cite{tbssrpphysica}. The present model, named
the extended effective exponential model or E$^3$M, considers bi-faceted impulsivity represented by two positive, random variables. This is a reasonable consideration given the fact that
impulsivity does fluctuate in a social system from person to person. We find that with this consideration, the subjective value of a reward follows a quasi power-law decay. However, unlike the EEM proposed in Ref.~\cite{tbssrpphysica}, this decay is now dictated by a special mathematical function called the confluent hypergeometric function of the second kind. 


\textcolor{black} {The novelty of the present paper can be elaborated based on two perspectives: Behavioural science and Physics. From the behavioural science perspective, the novelty of the paper lies in treating impulsivity as a multi-faceted quantity. Impulsivity is rarely mono-dimensional and existing mathematical models of delay discounting do not capture this feature. The proposed model is a better fit for realistic and complex real-world data leaving a scope for further generalization (e.g. introducing more facets). The delay discounting model is derived using superstatistics method that considers fluctuating parameters (impulsivity in delay discounting or temperature in a medium). Hence, our paper proposes a derivation of a novel mathematical model in behavioural science. We also observe is that although the two facets of impulsivity are often added for utility, addition may not be the only option, thus establishing a more nuanced role of the facets in behavioural science.
}

\textcolor{black} {
However, the importance of the model is not confined within behavioural science only.  There is a mathematical correspondence between temperature and impulsivity that has been utilized in Ref. \cite{tbssrpphysica} that proposes the effective exponential model with a mono-faceted impulsivity parameter. Also, just like impulsivity, temperature in a physical system too can have at least two facets.
For example, systems created in high-energy collisions (proton-proton, lead-lead, and so on) display a high-degree of anisotropy. Physics of high-energy collisions clearly distinguish between longitudinal (parallel to the incoming particle beam axis) and transverse (perpendicular to the incoming particle beam axis) dimensions. Realistic modelling describing physics of high-energy collision require two different temperatures in longitudinal and transverse direction. 
}

\textcolor{black} {
In physics of high-energy collisions, it can be shown that the mathematical form of the isotropic distributions describing particle production derived from nonadditive entropy can also be obtained from the superstatistics method utilized in this work. However, an anisotropic distribution following superstatistics approach has not been calculated in high-energy collisions yet. It would be interesting to compare the distributions calculated using the superstatistics method and the maximization of entropy method. Hence, our model in behavioural science bridges an inter-disciplinary gap because the mathematical technique proposed in the article may be utilized to model physics of high-energy collisions where two types of temperature will be relevant. 
}


As we proceed with the mathematical analysis, an additional constraint like $\qs<2$ is obtained. Given the constraint $\qs<2$, we argue that the 
$\langle \kr \rangle$ can be any positive number for observed datasets. However, from the datasets analyzed in this paper, we observe that $\langle \kr \rangle<1$. 
In this work, we consider datasets only from three different studies, and more datasets need to be analyzed with the help of the E$^3$M. \textcolor{black} {We also consider the two facets of impulsivity to be independent. However, it would be interesting to derive a generalized model considering a correlation between the two facets starting from a joint distribution of two correlated variables. We reserve this generalization for a future work.}

It is worth mentioning that although the present article studies only temporal discounting, the proposed formalism may be used when other types
of behaviour in effort discounting \cite{effortdisc}, probability discounting \cite{probdisc} and social discounting \cite{socdisc} are taken into account.

Another interesting formal development constitutes extending the E$^3$M, considering more facets of impulsivity. There are some behavioural studies relying on a three-factor \cite{bismod} (cognitive impulsivity, behavioural impulsivity, and impatience/restlessness), or even the four-factor UPPS model \cite{upps} (urgency, premeditation, perseverance, and sensation seeking) of impulsivity. We think that present method adapted in Section \ref{model} may very well be extended to models accommodating multiple ($>$2) variables, but we reserve that work for the future.



To conclude, the approach adapted in the article is a generalized one involving random variables. It is, by no means, limited only to the studies of impulsivity and delay discounting. In fact, any system represented by random variables has benefited (e.g., physical systems like a thermal plasma) and will benefit from these lines of argument. So, we hope that more and more investigations in this direction will throw light on the applicability of the work in explaining systems represented by multi-faceted random variables.


\vspace{20pt}
\section*{{ References}}
\bibliography{ref}  

\newpage
\centering \section*{{ Tables related to analysis of experimental data}}
\setlength{\arrayrulewidth}{0.1mm}
\setlength{\tabcolsep}{2pt}
\renewcommand{\arraystretch}{1.5}
\begin{table*}[h]
\centering 
\vspace{5pt}
\begin{tabular}{cccccc}
\hline
Study   & Reference                  &  Sample size                                         &      Age (years)                      & Gender                       & \hspace{-80pt} Other criteria     \\
\hline
\hline
1          
&  \cite{anokhin}
&   \makecell[l] {
560
(USA)}
& \makecell[c]{16-18
}

& \makecell[l]{Female 
\\= 50.7\%}
&  
\hspace{10pt}
\makecell[l]{
1. No head trauma \vspace{2pt} \\
2. No health conditions that restrict \\
\hspace{11pt}physical movement
}\\
\hline
2          
&  \cite{ordata}               
&                               		
 \makecell[l]{23 (Mexico)}
 &
  \makecell[c]{18-22 }
  &
  \makecell[l]
{Female \\= 65.2\%}	
& 
\hspace{10pt}
\makecell[l]{
1. No substance use problems \vspace{2pt} \\ 
2. No psychiatric diagnosis \vspace{2pt} \\
3. No psychiatric medication}		     
\\
\hline
3          
&  \cite{mansouridata}
&   
\makecell[l]{33 
(USA)}
& 
\makecell[c]{19-48 }
&   
 \makecell[l]
{
Female \\
= 57.6\%}
&
\hspace{10pt}
\makecell[l]{
1. No smokers \vspace{2pt} \\
2. No pregnant/breastfeeding participants \vspace{2pt} \\
3. No medication that affects appetite \vspace{2pt} \\
4. No allergies to study foods \vspace{2pt} \\
5. No active effort to lose weight 
}
\\
\hline
\end{tabular}
\caption{Descriptive statistics of the studies.}
\label{tabdesstat}
\end{table*}

\setlength{\arrayrulewidth}{0.1mm}
\setlength{\tabcolsep}{4pt}
\renewcommand{\arraystretch}{1.5}
\begin{table*}[h]
\centering
\vspace{0pt}
\begin{tabular}{cccccc}
\hline
Study   & Reference                  &  Dataset                                            &      $\qt$                      & $\qs$                       & $\langle \kr \rangle$     \\
\hline
\hline
1          &  \cite{anokhin}            &   Monozygotic, 16 yrs (M16)            & $6.463\pm0.816$    & $1.113\pm0.040$         & $0.023\pm0.010$ \\
            &                                    &   Monozygotic, 18 yrs (M18)            & $7.137\pm0.705$    & $1.134\pm0.038$        & $0.025\pm0.009$  \\
            &                                    &   \makecell[l]{Dizygotic, 16 yrs (D16)}                  & $5.414\pm0.800$    & $1.057\pm0.021$        & $0.014\pm0.006$  \\                            
            &                                    &   \makecell[l]{Dizygotic, 18 yrs (D18)}                  & $7.011\pm0.624$    & $1.120\pm0.030$        & $0.023\pm0.007$  \\                          
\hline
2          &  \cite{ordata}               &   \makecell[l]{Time point 1 (TP1)}                        &  $4.123\pm0.456$    & $1.145\pm0.044$        & $0.054\pm0.021$  \\      
	    &  	  			         &   \makecell[l]{Time point 2 (TP2)}                        &  $4.320\pm0.173$    & $1.287\pm0.011$       & $0.121\pm0.009$  \\ 
	    &                 			&   \makecell[l]{Time point 3 (TP3)}                        &   $4.443\pm0.450$    & $1.319\pm0.028$       & $0.136\pm0.025$  \\                               
\hline
3          &  \cite{mansouridata}   &               	-				&   $8.893\pm0.397$    &	$1.313\pm0.076$     & $0.058\pm0.020$   \\
\hline
\end{tabular}
\caption{A summary of the results obtained from description of existing datasets using the E$^3$M.}
\label{tabsum}
\end{table*}

\newpage 
\centering \section*{{ Figures related to analysis of experimental data}}

\vspace{30pt} 
 
\subsection*{1. Study 1} 

\newpage
\subsection*{2. Study 2}

\newpage

\subsection*{3. Study 3}


\subsection*{4. Comparison of the results obtained from all the datasets used in the paper}
 
\end{document}